\documentclass[aps,twocolumn,prl,showpacs,floatfix,superscriptaddress]{revtex4}

\usepackage{amsmath,fixmath}
\usepackage{graphicx}
\usepackage{dcolumn}
\usepackage{amssymb}
\usepackage{color}
\usepackage{natbib}
\usepackage{hyperref}

\newcommand{\vn}[1]{\boldsymbol #1}
\newcommand{\tw}{\ensuremath{t_\mathrm{w}}}

\begin{document}

\title{Static versus dynamic heterogeneities in the \boldmath $D\!=\!3$
Edwards-Anderson-Ising spin glass}

\author{R.~Alvarez~Ba\~nos} \affiliation{Instituto de Biocomputaci\'on y
  F\'{\i}sica de Sistemas Complejos (BIFI), 50009 Zaragoza, Spain.}
  \affiliation{Departamento
  de F\'\i{}sica Te\'orica, Universidad
  de Zaragoza, 50009 Zaragoza, Spain.} 

\author{A.~Cruz} \affiliation{Departamento
  de F\'\i{}sica Te\'orica, Universidad
  de Zaragoza, 50009 Zaragoza, Spain.} \affiliation{Instituto de Biocomputaci\'on y
  F\'{\i}sica de Sistemas Complejos (BIFI), 50009 Zaragoza, Spain.}

\author{L.A.~Fernandez} \affiliation{Departamento
  de F\'\i{}sica Te\'orica I, Universidad
  Complutense, 28040 Madrid, Spain.} \affiliation{Instituto de Biocomputaci\'on y
  F\'{\i}sica de Sistemas Complejos (BIFI), 50009 Zaragoza, Spain.}

\author{J.~M.~Gil-Narvion} \affiliation{Instituto de Biocomputaci\'on y
  F\'{\i}sica de Sistemas Complejos (BIFI), 50009 Zaragoza, Spain.}

\author{A.~Gordillo-Guerrero}\affiliation{D. de  Ingenier\'{\i}a
El\'ectrica, Electr\'onica y Autom\'atica, U. de Extremadura,
  10071, C\'aceres, Spain.}\affiliation{Instituto de Biocomputaci\'on y
  F\'{\i}sica de Sistemas Complejos (BIFI), 50009 Zaragoza, Spain.}

\author{M.~Guidetti} \affiliation{Dipartimento
  di Fisica, Universit\`a di Ferrara and INFN - Sezione di Ferrara,
  Ferrara, Italy.} 

\author{A.~Maiorano} 
  \affiliation{Dipartimento di Fisica, INFM and
  INFN, Universit\`a di Roma ``La Sapienza'', 00185 Roma, Italy.}\affiliation{Instituto de Biocomputaci\'on y
  F\'{\i}sica de Sistemas Complejos (BIFI), 50009 Zaragoza, Spain.}

\author{F.~Mantovani} \affiliation{DESY, D-15738 Zeuthen, Germany.}

\author{E.~Marinari} \affiliation{Dipartimento di Fisica, INFM and
  INFN, Universit\`a di Roma ``La Sapienza'', 00185 Roma, Italy.}

\author{V.~Martin-Mayor} \affiliation{Departamento de F\'\i{}sica
  Te\'orica I, Universidad Complutense, 28040 Madrid, Spain.} \affiliation{Instituto de Biocomputaci\'on y
  F\'{\i}sica de Sistemas Complejos (BIFI), 50009 Zaragoza, Spain.}

\author{J.~Monforte-Garcia} \affiliation{Instituto de Biocomputaci\'on y
  F\'{\i}sica de Sistemas Complejos (BIFI), 50009 Zaragoza, Spain.}
  \affiliation{Departamento
  de F\'\i{}sica Te\'orica, Universidad
  de Zaragoza, 50009 Zaragoza, Spain.}

\author{A.~Mu\~noz Sudupe} \affiliation{Departamento
  de F\'\i{}sica Te\'orica I, Universidad
  Complutense, 28040 Madrid, Spain.} 

\author{D.~Navarro} \affiliation{D.  de Ingenier\'{\i}a,
  Electr\'onica y Comunicaciones and I3A, U. de Zaragoza, 50018 Zaragoza, Spain.}

\author{G.~Parisi} \affiliation{Dipartimento di Fisica, INFM and
  INFN, Universit\`a di Roma ``La Sapienza'', 00185 Roma, Italy.}

\author{S.~Perez-Gaviro} \affiliation{Instituto de Biocomputaci\'on y
  F\'{\i}sica de Sistemas Complejos (BIFI), 50009 Zaragoza, Spain.}
 \affiliation{Dipartimento di Fisica, INFM and
  INFN, Universit\`a di Roma ``La Sapienza'', 00185 Roma, Italy.}

\author{J.~J.~Ruiz-Lorenzo} \affiliation{Departamento de
  F\'{\i}sica, Universidad de Extremadura, 06071 Badajoz, Spain.}\affiliation{Instituto de Biocomputaci\'on y
  F\'{\i}sica de Sistemas Complejos (BIFI), 50009 Zaragoza, Spain.}

\author{S.F.~Schifano} \affiliation{Dipartimento
  di Fisica Universit\`a di Ferrara and INFN - Sezione di Ferrara,
  Ferrara, Italy.} 

\author{B.~Seoane} \affiliation{Departamento de F\'\i{}sica
  Te\'orica I, Universidad Complutense, 28040 Madrid, Spain.} \affiliation{Instituto de Biocomputaci\'on y
  F\'{\i}sica de Sistemas Complejos (BIFI), 50009 Zaragoza, Spain.}

\author{A.~Tarancon} \affiliation{Departamento
  de F\'\i{}sica Te\'orica, Universidad
  de Zaragoza, 50009 Zaragoza, Spain.} \affiliation{Instituto de Biocomputaci\'on y
  F\'{\i}sica de Sistemas Complejos (BIFI), 50009 Zaragoza, Spain.}

\author{R.~Tripiccione} \affiliation{Dipartimento
  di Fisica Universit\`a di Ferrara and INFN - Sezione di Ferrara,
  Ferrara, Italy.} 
 
\author{D.~Yllanes}  \affiliation{Departamento de F\'\i{}sica
  Te\'orica I, Universidad Complutense, 28040 Madrid, Spain.}\affiliation{Instituto de Biocomputaci\'on y
  F\'{\i}sica de Sistemas Complejos (BIFI), 50009 Zaragoza, Spain.}

\date{\today}

\begin{abstract}
We numerically study the aging
properties of the dynamical heterogeneities in the Ising spin
glass. We find that a phase transition takes place during the aging
process. Statics-dynamics correspondence implies that systems of
finite size in equilibrium have static heterogeneities that obey 
Finite-Size Scaling, thus signaling an analogous phase transition
in the thermodynamical limit. We compute the critical exponents and the
transition point in the equilibrium setting, and use them to show that
aging in dynamic heterogeneities can be described by a
Finite-Time Scaling Ansatz, with potential implications
for experimental work.
\end{abstract}
\pacs{75.50.Lk, 75.40.Mg, 75.10.Nr} 
\maketitle

Spin glasses, fragile molecular glasses, polymers, colloids, and many
other materials display a dramatic increase of characteristic times
when cooled down to their glass temperature,
$T_\mathrm{g}$~\cite{cavagna:09}. This
is probably due to the collective movements of an
increasing number of elements in the system, with a (free) energy
barrier growing with the size of the cooperative
regions~\cite{adam:65} (the cooperative regions become larger
as the temperature gets closer to $T_\mathrm{g}$).
Experimentally, one can get the
fingerprints of these movements by observing dynamical
heterogeneities~\cite{weeks:00} or non-linear
susceptibilities~\cite{berthier:05}.

Below $T_\mathrm{g}$, {\em aging} appears~\cite{vincent:97}. Consider
a rapid quench from a high temperature to the working
temperature $T$ ($T<T_\mathrm{g}$), where the system is left to equilibrate
for time $t_\mathrm{w}$ and probed at a later time $t+t_\mathrm{w}$.
One finds that the response functions (e.g., magnetic susceptibility)
depend on $t/t_\mathrm{w}^\mu$, with $\mu\approx
1$~\cite{vincent:97,rodriguez:03,dupuis:05}.
The age of the glass, $t_\mathrm{w}$,
remains the relevant time scale even for
$t_\mathrm{w}\sim $ days. 

Dynamical heterogeneities age as well, as found numerically in their
characteristic length
$\zeta(t,t_\mathrm{w})$~\cite{jaubert:07,janus:09b}. Recent
measurements of aging correlation and response functions with
space-time resolution~\cite{oukris:10} suggest that
$\zeta(t,t_\mathrm{w})$ will soon be experimentally investigated.
Characterizing aging for $\zeta(t,t_\mathrm{w})$ is our main concern
here.

We focus on spin glasses, an easier case for a number of
reasons: (i) the sluggish dynamics is known to be due to a
thermodynamic phase transition at
$T_\mathrm{c}=T_\mathrm{g}$~\cite{gunnarsson:91,palassini:99,ballesteros:00};
(ii) the size of the {\em glassy} magnetic domains,
$\xi(t_\mathrm{w})$, is experimentally
accessible~\cite{joh:99,bert:04} ($\xi\sim 100$ lattice spacings at
$T\sim T_\mathrm{c}$~\cite{joh:99}, larger than comparable
measurements for structural glasses~\cite{berthier:05}); (iii)
$\xi(t_\mathrm{w})\propto t_\mathrm{w}^{1/z(T)}$, $z(T)\approx
6.9 T_\mathrm{c}/T$~\cite{janus:09b} suggests that
free-energy barriers grow in spin glasses as $\sim
\log\xi(t_\mathrm{w})$, rather than with a power law as in fragile
glasses; (iv) equilibrium physics is known to rule nonequilibrium
dynamics~\cite{franz:98}. A quantitative correspondence exists between
equilibrium and nonequilibrium spatial correlation
functions~\cite{janus:08b,janus:10} (equilibrium on systems of {\em
  size} $L$ matches nonequilibrium at {\em time}
$t_\mathrm{w}$). Finally, the Janus dedicated computer~\cite{janus:06}
allows us to simulate nonequilibrium dynamics from picoseconds to a tenth
of a second~\cite{janus:08b,janus:09b}, and to compute equilibrium
correlation functions on lattices as large as 
$L=32$, and temperatures as low as
$0.64 ~T_\mathrm{c}$~\cite{janus:10}.

In this paper we show that a phase transition occurs in the aging dynamic
heterogeneities. As time $t$ proceeds, when the spin correlation
function $C(t,t_\mathrm{w})$ (see below) becomes smaller than the spin
glass order parameter $q_\mathrm{EA}$, the length scale of the dynamic
heterogeneities $\zeta(t,t_\mathrm{w})$ diverges in the limit of large
$t_\mathrm{w}$. We use the statics-dynamics correspondence to
investigate this phase transition in the equilibrium setting,
focusing on spatial correlation functions ({\em static}
heterogeneities). Finite-Size Scaling (FSS) yields an accurate
estimate of $q_\mathrm{EA}$ (something never achieved before for a 
spin glass) as well as of the relevant critical exponents. Back to
nonequilibrium, aging turns out to be amazingly well described by a
{\em Finite-Time Scaling Ansatz}, with critical parameters taken
verbatim from the equilibrium computation.

We consider the Edwards-Anderson model on a cubic lattice of size $L$
(volume $V=L^3$), with periodic boundary conditions, at $T=0.64
T_\mathrm{c}$. We use Ising spins, $s_{\vn x}=\pm1$, and binary
nearest-neighbor couplings.  
The average over the quenched disorder, denoted by an
overline, is taken after the thermal average
$\langle\ldots\rangle$.  We consider two {\em clones} of the system,
$\{s_{\vn{x}}^{(1)},s_{\vn{x}}^{(2)}\}$ evolving independently under
the same set of coupling constants, and taken at the same time
$t_\mathrm{w}$. The {\em replica} field is
$q_{\vn{x}}=s_{\vn{x}}^{(1)}s_{\vn{x}}^{(2)}$ 
and the spin
overlap is its spatial average $q=\sum_{\vn{x}} q_{\vn{x}}/V$.
See Refs.~\cite{janus:10}
and~\cite{janus:09b} for full details of our equilibrium and nonequilibrium
simulations.

\begin{figure}
\includegraphics[height=\linewidth,angle=270]{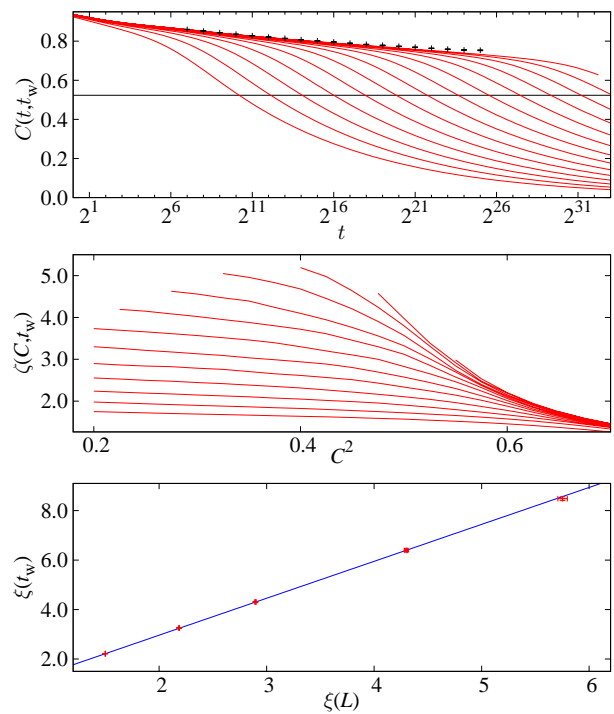}
\caption{(color online) Top: $C(t,\tw)$, Eq.~\eqref{eq:Cttw}, as a
  function of $t$ for $\tw=4^i$, $i=3,..,16$ (lines, $\tw$ grows from
  bottom to top). We also plot $\tilde C(t)$ (points) and our result
  for $q_\mathrm{EA}$ from Eq.~\eqref{eq:qEA} (horizontal
  line). Center: Correlation length $\zeta(C,\tw)$ as a function of
  $C^2$ (same values of $\tw$ as in the top panel). Bottom:
  Finite-time coherence length $\xi(\tw)$ against the finite-size
  coherence length at $q=0$, $\xi(L)$. The results are compatible with
  $\xi(\tw) = 1.48\xi(L)$ (straight line).  }\label{fig:Introduccion}
\end{figure}

Out of equilibrium, correlation functions depend either on a
single time $t_\mathrm{w}$, or on $t$ and $t_\mathrm{w}$. Let
$c_{\vn{x}}(t,t_\mathrm{w})=s_{\vn{x}}(t+t_\mathrm{w})
s_{\vn{x}}(t_\mathrm{w})$.  The spin correlator, see
Fig.~\ref{fig:Introduccion}--top, is
\begin{equation}\label{eq:Cttw}
C(t,t_\mathrm{w})=\frac{1}{V} \sum_{\vn{x}} \overline{\langle c_{\vn{x}}(t,t_\mathrm{w})\rangle}\  ,\ \tilde C(t)= C(t,t_\mathrm{w}=\infty)\,. 
\end{equation}
Naive aging is approximatively valid: for finite $t_\mathrm{w}$,
$C(t,t_\mathrm{w})$ decays for long $t$, but the decay slows down with
increasing $\tw$. In fact, there is an enveloping
curve $\tilde C(t)$ with a non-zero limiting value, the order
parameter $q_\mathrm{EA}$. The lack of a reliable parameterization
of $\tilde C(t)$  precludes a controlled extrapolation of $q_\text{EA}$,
in contrast with the equilibrium computation shown below.

As for space dependencies, we consider
$C_4(\vn{r},t_\mathrm{w})=\sum_{\vn{x}} \overline{\langle
  q_{\vn{x}}(t_\mathrm{w})q_{\vn{x}+\vn{r}}(t_\mathrm{w})\rangle}/V$.
Using integral estimators~\cite{janus:08b,janus:09b} we extract
the coherence length $\xi(t_\mathrm{w})$, the size of regions where
the two clones of the system are similar. Yet, to learn about
heterogeneities on the dynamics probed at time $t+t_\mathrm{w}$, at
distance $\vn{r}$, we consider $C_{22}(\vn{r},t,t_\mathrm{w})=
\sum_{\vn{x}} \overline{\langle c_{\vn{x}}(t,t_\mathrm{w})
  c_{\vn{x}+\vn{r}}(t,t_\mathrm{w})\rangle -
  C^2(t,t_\mathrm{w})}/V\,.$ Using an integral
estimator~\cite{janus:09b}, we extract from $C_{22}$ the
correlation-length $\zeta(t,t_\mathrm{w})$, the characteristic length
for heterogeneities, displayed in the central panel of
Fig.~\ref{fig:Introduccion}. 
We replace $t$ with $C(t,t_\mathrm{w})$~\cite{cugliandolo:93}, as independent variable. 
For large $C$,
$\zeta(C,t_\mathrm{w})$ reaches a $t_\mathrm{w}$-independent value,
which increases when $C$ decreases. On the other hand, for small $C$,
$\zeta(C,t_\mathrm{w})$ grows strongly with $t_\mathrm{w}$. Clearly,
something happens when $C$ goes through some special value
$q_\mathrm{EA}$ and we intend to exploit the statics-dynamics
correspondence to clarify it.

How does all this appear from an equilibrium viewpoint?  In the limit of
large system size $L$, the probability density function for $q$,
$P(q)=\overline{P_J(q)}$, has two Dirac's delta contributions of equal
weight at $q=\pm q_\mathrm{EA}$. Replica Symmetry Breaking (RSB)
theory~\cite{marinari:00} predicts that $P(q)$ has a support for
$|q|<q_\mathrm{EA}$, while droplet theory expects no support in that
region~\cite{bray:87}. 

Our approach focuses on the study of equilibrium {\em connected}
correlation functions~\cite{contucci:09}, regarded as a function of
the spin overlap $q$. Varying $q$ at fixed $T$ a phase
transition is encountered for $q\!=\!q_\mathrm{EA}$.  As
in~\cite{janus:10}, our conditional correlation function at fixed
$q=c$, $C_4(\vn r|c)$, is obtained as a quotient of the convolutions
of $\overline{\langle q_{\vn x} q_{\vn x + \vn r} \delta(q-c)\rangle}$
and $\overline{\langle \delta(q-c)\rangle}$
with a Gaussian of width $1/V$. This combines $\mathcal O(\sqrt V)$ levels,
thus smoothing the comb-like $P(q)$~\cite{fernandez:09}. 

It has been recently found~\cite{janus:08b,janus:10} that the {\em
  equilibrium} $C_4(\vn{r}|q)$, computed in a system of size $L$,
accurately matches the nonequilibrium $C_{22}({\vn r},C,t_\mathrm{w})$
if one chooses time $t$ such that $C(t,t_\mathrm{w})=q$ and time
$t_\mathrm{w}$ such that $L\approx 3.7\xi(t_\mathrm{w})$ (at least at
$T=0.64 T_\mathrm{c}$). It is tempting to assume that the
correspondence will become exact in the limit of large $L$ and
$t_\mathrm{w}$.  In Fig.~\ref{fig:Introduccion}--bottom we show an
example of this correspondence in the limit $C\to 0$.

To proceed with the equilibrium analysis, we observe that
$C_4(\vn{r}|q)$ tends to $q^2$ for large $r$. In a finite system, one
needs to perform a subtraction that complicates the analysis
\cite{contucci:09}. We instead consider the Fourier transform at wave
vector $\vn{k}$, $\hat C_4(\vn{k}|q)=
\sum_{\vn{r}}\,\mathrm{e}^{\mathrm{i}\vn{k}\cdot\vn{r}}
C_4(\vn{r}|q)$, blind to a constant subtraction for $\vn k\neq0$. 
Defining $\vn k_\text{min}\!=\! (2\pi/L,0,0)$ (or
permutations), we have
\begin{align}\label{eq:F}
F_q &= \hat C_4\bigl.\bigl(\vn k_\text{min} \bigr|\, q\bigr).
\end{align}

For $T<T_\mathrm{c}$ and $|q|\leq q_\mathrm{EA}$, one expects that
\begin{align}
C_4(\vn{r}|q)&\simeq q^2 + \frac{A_q}{r^{\theta(q)}},&
\hat C_4(\vn{k}|q)&\propto
k^{\theta(q)-D}+\ldots\label{EQ:SCALINGC4Q-LARGE-L-SMALL-Q}
\end{align}
(scaling in Fourier space holds only if $\theta(q)<D$).  The dots
in~\eqref{EQ:SCALINGC4Q-LARGE-L-SMALL-Q} stand for scaling
corrections, subleading in the limit of large $r$ (or small $k$). On
the other hand~\cite{contucci:09},
\begin{equation}
C_4(\vn{k}\,|\,q^2>q_\mathrm{EA}^2) \propto \frac{1}{k^2 +\ \xi^{-2}_q}\,.\label{EQ:SCALINGC4Q-LARGE-L-LARGE-Q}
\end{equation}
The correlation length $\xi_q$ diverges when $|q|\to q_\mathrm{EA}$
from {\em above}, $\xi_q^{L=\infty}\!\propto\!
(q^2-q^2_\mathrm{EA})^{-\hat\nu}$. In principle, $\hat\nu$ is
different from the thermal critical exponent at $T_\mathrm{c}$.  We
note as well the scaling law~\footnote{If we add an interaction
  $h\,q\,L^{D}$ to the Hamiltonian, (i) the correlation length
  diverges as $\xi(h)\!\propto\!h^{-\nu_\mathrm{h}}$; (ii) $h\propto
  [q_\mathrm{EA}(h)-q_\mathrm{EA}(0)]^{1/[1-\nu_\mathrm{h}(D-\theta(q_\mathrm{EA})]}$,
  since $\mathrm{d} q_\mathrm{EA}/{\mathrm d} h \!\propto\!
  \xi^{D-\theta(q_\mathrm{EA})}$; and (iii)
  $\theta(q_\mathrm{EA})\!=\!2(D -\nu_\mathrm{h}^{-1})$ (see,
  e.g.,~\cite{amit:05}). Observe that
  $\nu^{-1}_\mathrm{h}+\hat\nu^{-1}\!=\!D$.}
\begin{equation}
\theta(q_\mathrm{EA})=2/\hat\nu\,.\label{eq:ley-de-hiperescala}
\end{equation}

The theories for the spin-glass phase differ in the precise form of $\theta(q)$,
but agree that a crossover can be
detected in $F_q$ for finite $L$.  Indeed, $F_q\sim L^{D-\theta(q)}$
for $|q| <q_\mathrm{EA}$, while $F_q\sim 1$ for $|q|> q_\mathrm{EA}$,
see
Eqs.~(\ref{EQ:SCALINGC4Q-LARGE-L-SMALL-Q},\ref{EQ:SCALINGC4Q-LARGE-L-LARGE-Q}).
For large $L$, $\xi_q^{L=\infty}\!\propto\!
(q^2-q^2_\mathrm{EA})^{-\hat\nu}$ and the crossover becomes a phase
transition.  FSS tells us that, see e.g.~\cite{amit:05},
\begin{equation}
F_q = L^{D-\theta(q_\mathrm{EA})} G\bigl(L^{1/\hat\nu} (q-q_\mathrm{EA})\bigr)\,,\label{FSSA1}
\end{equation}
up to scaling corrections.  $G$ is a scaling function. 

\begin{figure}
\includegraphics[height=\linewidth,angle=270]{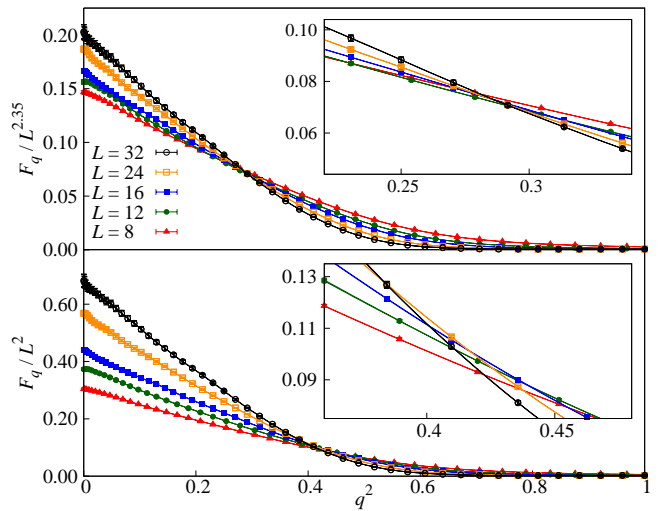}
\caption{(color online) $F_q/L^y$, Eq.~\eqref{eq:Fy}, against $q^2$ for all
  our $L$ values at $T=0.703\approx 0.64 T_\mathrm{c}$, both for
  the free-field scaling ($F_q(\vn{k})\propto 1/k^2$, i.e., $y=2$) and for
  $y=2.35\approx D-\theta(q_\mathrm{EA})$.
  The insets are closeups of
  the crossing region.}\label{fig:cortes}
\end{figure}

We have exploited Eq.~\eqref{FSSA1} in the following non-standard 
way. We focus
 on quantities depending on the continuous parameter
$y$ ($\epsilon=D-\theta(q_\mathrm{EA})$):
\begin{align}\label{eq:Fy}
F_q/L^y&= L^{\epsilon-y} G\bigl(L^{1/\hat\nu} (q-q_\mathrm{EA})\bigr)\,. 
\end{align}
When $y$ is smaller than $D-\theta(0)$, $F_q/L^y$
vanishes in the large-$L$ limit for $|q|> q_\mathrm{EA}$, while it
diverges for $|q| < q_\mathrm{EA}$. Hence, fixing $y$,
the curves for  pairs of lattices $(L,2L)$, will
cross at a point $q_{L,y}$, see Fig.~\ref{fig:cortes}.
To leading order in $L^{-1}$, the crossing point 
approaches $q_\mathrm{EA}$ for large $L$ as
\begin{equation}
q_{L,y}=q_\mathrm{EA} + A_{y}
L^{-1/\hat\nu}\,,\ A_{y}=\frac{G(0)}{G'(0)}\frac{2^{\epsilon-y}-1}{2^{1/\hat\nu}-2^{\epsilon-y}}\,.\label{eq:cortes}
\end{equation}
Note that  the amplitude $A_y$ changes
sign at $y\!=\!\epsilon$. 

\begin{figure}
\includegraphics[height=\linewidth,angle=270]{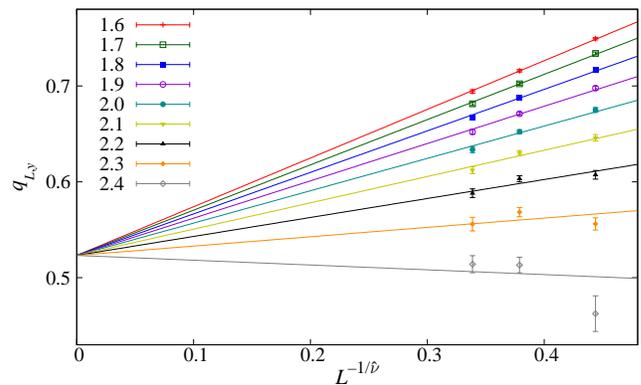}
\caption{(color online) Crossing points $q_{L,y}$ for $F_q/L^y$ 
  computed from pairs of lattices $(L,2L)$, versus $L^{-1/\hat\nu}$, for different $y$
  values at $T\!=\!0.703$. The continuous lines are fits to Eq.~(\ref{eq:cortes}),
  constrained to yield common values for $q_\mathrm{EA}$ and
  $1/\hat\nu$.}\label{fig:ajuste-cortes}
\end{figure}
\begin{figure}
\centering \includegraphics[height=\linewidth,angle=270]{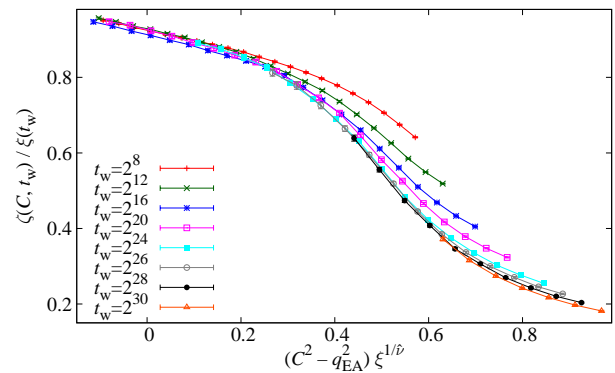}
\caption{(color online) Dimensionless ratio $\zeta(C,\tw)/\xi(\tw)$ 
against the scaling variable $(C^2-q_\mathrm{EA}^2) \xi(\tw)^{1/\hat\nu}$.
}
\label{fig:scaling_dinamica}
\end{figure}

We could use a fit to Eq.~\eqref{eq:cortes} in order to obtain the
order parameter, but, for a fixed $y$, there are only three crossing
points ($L=8,12,16$) for three parameters (zero degrees of freedom).
Fortunately, one can extract more information from the data by
computing the crossings for several $y$ values and performing a joint
fit, sharing $q_\mathrm{EA}$ and $1/\hat\nu$.  Since
these additional crossing points are not statistically independent,
this procedure requires a proper consideration of the
cross-correlations. This can be achieved by computing the fit goodness
estimator $\chi^2$ with the full covariance matrix for the
$q_{L,y}$. The number of $y$ values considered is a compromise between
adding more degrees of freedom and keeping the covariance matrix
invertible. We have chosen 9 values of $y$ obtaining $\chi^2=18.9$,
reasonable for a fit with $16$ degrees of freedom (see
Fig.~\ref{fig:ajuste-cortes}).  The result is
\begin{equation}\label{eq:qEA}
q_\mathrm{EA}=0.52(3)\,,\quad 1/\hat\nu=0.39(5)\,.
\end{equation}
These numbers are remarkably stable to variations in the 
set of $y$ values.  
Also, removing the $L=8$ data for $y=2.3,2.4$ (the outliers in
Fig.~\ref{fig:ajuste-cortes}) shifts our results by one fifth of the error
bars.
Note as well that the slope $A_{y}$ changes sign at
$y\approx 2.35$. Hence, $\theta(q_\mathrm{EA})\approx 0.65$,
in reasonable agreement with Eq.~(\ref{eq:ley-de-hiperescala}).  

The value of $q_\mathrm{EA}$ computed above should be the same as
the large-$L$ extrapolation of the position of the peak in $P (q)$: a fit
$q_\mathrm{EA}(L)=q_\mathrm{EA} + A L^{-1/\hat\nu}$, with $\hat\nu$ from
\eqref{eq:qEA}, yields
$q_\mathrm{EA}=0.54(3)$ \cite{janus:10}. 

We are finally ready to discuss aging in the dynamic
heterogeneities. The statics-dynamics correspondence suggests that it
will take the form of a {\em Finite-Time Scaling} (FTS) Ansatz, similar to
Eq.~(\ref{FSSA1}), in which $\xi(t_\mathrm{w})$ plays the role of $L$.
Since it is a length, $\zeta(C,t_\mathrm{w})$ should have the same
scaling dimensions of $\xi(t_\mathrm{w})$.  Setting short-time
corrections aside, Fig.~\ref{fig:scaling_dinamica} shows indeed that
$\zeta(C,t_\mathrm{w})/\xi(t_\mathrm{w})$ behaves as a function of
$(C^2-q_\mathrm{EA}^2)\xi(t_\mathrm{w})^{1/\hat\nu}$.
FTS also provides a natural explanation for the extremely small
exponents found in $t$-extrapolations for $\tilde C(t)$ \footnote{
In \cite{janus:09b}, we find $\tilde C(t)-q_\mathrm{EA}\propto t^{-a}$ with
$a\approx 0.05$.  Clearly enough, FTS implies that
$\tilde C(t)-q_\mathrm{EA}\propto \tilde \zeta(t)^{-1/\hat\nu}$
[$\tilde \zeta(t)\!=\!\zeta(t,t_\mathrm{w}\!=\!\infty)$]. Now, full-aging (as
well as empirical evidence~\cite{janus:09b}) suggest that $\tilde
\zeta(t)\propto t^{T/[T_\mathrm{c} z(T_\mathrm{c})]}$, just as
$\xi(t_\mathrm{w})$. Therefore $a\!=\!T/[T_\mathrm{c}
  \hat\nu z(T_\mathrm{c})]\approx 0.04$.}.

In summary, we have studied aging properties in glassy dynamic
heterogeneities for the Ising spin glass,
 characterized through their characteristic length
$\zeta(t,t_\mathrm{w})$. Aging takes the
form of a Finite-Time Scaling Ansatz, which describes the crossover
from a regime where $\zeta(t,t_\mathrm{w})$ is of order one, to a
regime where it is of order $\xi(t_\mathrm{w})$, the coherence length
yielding the size of the glassy magnetic domains. In the limit of an
infinite waiting time, the crossover evolves into a phase
transition. We have profited from the statics-dynamics
correspondence~\cite{janus:08b,janus:10} to study this phase
transition via {\em equilibrium} spatial correlation functions, thus
obtaining the critical exponents and, for the first time, the
spin-glass order parameter. These critical parameters, taken verbatim,
describe our nonequilibrium data. For a discussion of
the mode-coupling transition in glass-forming liquids in a similar vein
see~\cite{franz:10}.  

Janus was supported by EU FEDER funds (UNZA05-33-003, MEC-DGA, Spain),
and developed in collaboration with ETHlab. We were partially
supported by MICINN (Spain), through contracts No. TEC2007-64188,
FIS2007-60977, FIS2009-12648-C03, by Junta de Extremadura (GRU09038),
and by UCM-Banco de Santander.  BS and DY were supported by the FPU
program (Spain) and SPG by FECYT (Spain).

\end{document}